\documentclass{aa}
\usepackage{graphicx}
\usepackage{txfonts}
\usepackage{natbib}
\bibpunct{(}{)}{;}{a}{}{,}
\begin{document} 
   \title{Pleiades low-mass brown dwarfs: the cluster L dwarf sequence}
%     \thanks{}
%
   \titlerunning{Pleiades low-mass brown dwarfs}
   \author{G.~Bihain\inst{1}\fnmsep\inst{2}
      \and R.~Rebolo\inst{1}\fnmsep\inst{2}
      \and V.~J.~S.~B\'ejar\inst{1}\fnmsep\inst{3}
      \and J.~A.~Caballero\inst{1}
      \and C.~A.~L.~Bailer-Jones\inst{4}
      \and R.~Mundt\inst{4}
      \and J.~A.~Acosta-Pulido\inst{1}
      \and A.~Manchado~Torres\inst{1}\fnmsep\inst{2}}
   \authorrunning{G. Bihain et al.}
   \institute{}
   \institute{Instituto de Astrof\'{\i}sica de Canarias, c/ V\'{\i}a
	      L\'actea,
	      s/n, 38205  La Laguna, Tenerife, Spain\\
%	      \email{gbihain@ll.iac.es}
	      \email{[gbihain,rrl,zvezda,jap,amt]@ll.iac.es,victor.bejar@gtc.iac.es}
	      \and
	      Consejo Superior de Investigaciones Cient\'{\i}ficas, Spain
	      \and
	      GTC Project, Instituto de Astrof\'{\i}sica de Canarias
	      \and
	      Max-Planck-Institut f{\"u}r Astronomie, K{\"o}nigstuhl 17,
	      69117 Heidelberg, Germany\\
	      \email{[calj,mundt]@mpia-hd.mpg.de}
	     }
   
   \date{Received date; accepted date} 

   \abstract
   {}  
   {We present a search for low-mass brown dwarfs in the 
Pleiades open cluster. The identification of Pleiades members
fainter and cooler than those currently known allows us to
constrain evolutionary models for L dwarfs and to extend the
study of the cluster mass function to lower masses.}
      {We conducted a 1.8~deg$^2$ near-infrared $J$-band survey at
the 3.5~m Calar Alto Telescope, with completeness
$J_{cpl}\sim$~19.0. The detected sources were correlated with
those of previously available optical $I$-band images
($I_{cpl}\sim$~22). Using a $J$~versus~$I-J$ colour--magnitude
diagram, we identified 18 faint red L-type candidates, with
magnitudes $17.4<J<$~19.7 and colours $I-J>$~3.2. If Pleiades
members, their masses would span $\sim$0.040--0.020~$M_{\sun}$.
We performed follow-up $HK_{\rm s}$-band imaging to further confirm
their cluster membership by photometry and proper motion.}
     {Out of 11 $IJ$ candidates with proper motion measurements, we find six
cluster members, two non-members and three whose membership is
uncertain and depends on the intrinsic velocity dispersion of
Pleiades brown dwarfs. This dispersion ($>$4~mas~yr$^{-1}$) is at
least four times that of cluster stars with masses
$\ga$1~$M_{\sun}$. Five of the seven other $IJ$ candidates are
discarded because their $J-K_{\rm s}$ colours are bluer than those
of confirmed members. Our least massive proper motion members are
\object{BRB~28} and 29 ($\sim$25~$M_{Jup}$). The $J$~versus~$I-J$
sequence of the L-type candidates at $J>18$ is not as red as
theoretical models predict; it rather follows the field L-dwarf
sequence translated to the cluster distance. This sequence
overlapping, also observed in the $J$~versus~$J-H$ and $J-K$
diagrams, suggests that Pleiades and field L dwarfs may have
similar spectral energy distributions and luminosities, and thus
possibly similar radii. Also, we find $\alpha=$~$0.5\pm0.2$ for a power-law
approximation $dN/dM \propto M^{-\alpha}$ of the survey mass
spectrum in the mass range 0.5--0.026~$M_{\sun}$. This value is similar to
that of much younger clusters, indicating no
significant differential evaporation of low-mass Pleiades members
relative to more massive ones.} 
% conclusions heading (optional)
  {}


\maketitle

\section{Introduction}

As a result of progressive cooling, brown dwarfs are expected to evolve from
late M~spectral type at very early stages (age $\la$~10~Myr) to L-type ($T_{\rm
eff}$ in the range $\sim$2\,400 to $\sim$1\,400~K; \citealt{dahn2002}), and at
sufficiently old ages (a few Gyr), to T-type and beyond ($T_{\rm eff}$
$\la$~1\,400~K). The mass-spectral type and mass-luminosity relationships at
different ages remain to be derived from observations of cool dwarfs, which
also permit the calibration of the theory of substellar evolution (e.g.
\citealt{chabrier2000a}). In comparison to the spectra of M dwarfs, L dwarfs
are characterized in the optical by the weakening of the metal-oxide TiO and VO
bands due to the condensation of the metals Ti and V in dust grains that may
significantly affect the atmospheric structure and the emergent spectrum (e.g.
the DUSTY models from \citealt{chabrier2000b}). Bands of metal-hydrides CrH,
FeH and CaH and bands of water become more intense especially in the
near-infrared (IR), whereas the neutral alkali metals Na, K, Rb, Cs and Li
become stronger in the optical \citep{chabrier2000a,basri2000}. The emergence
in the near-IR of CH$_{4}$ bands characterizes the beginning of the T dwarf
sequence. Field L dwarfs have $I-J$ and $J-K_{\rm s}$ colours redder than
$\sim$3.3 and $\sim$1.3~mag, respectively, increasing from early to later
spectral subclasses \citep{martin1999}. For the L4--L8 spectral subclasses the
near-IR $J-H$ and $J-K_{\rm s}$ colours appear to saturate at $\sim$1.2 and
$\sim$2.0~mag \citep{kirkpatrick2000,burgasser2002}, respectively, whereas for
T dwarfs these colours become bluer, with values decreasing to below zero
\citep{knapp2004}. From L0 to T8 spectral types the $I-J$ colour is increasing
to $\sim$5.8~mag \citep{dahn2002}. Associated with the other colours it can
give a rough indication of the spectral type.

The \object{Pleiades} open cluster ($\sim$120~Myr, $\sim$130~pc)
served for decades as a reference stellar laboratory where models
were contrasted with observations. Since the discovery of Pleiades
brown dwarfs \citep{rebolo1995,rebolo1996}, subsequent studies have
provided numerous fainter substellar candidates with spectral types
down to late M
(\citealt{zapateroosorio1997a,zapateroosorio1997b,bouvier1998},
hereafter\defcitealias{bouvier1998}{B98}B98;
\citealt{festin1998a,zapateroosorio1999Pl}), from which a fraction
is confirmed by lithium detection \citep{stauffer1998a,martin2000}
or by proper motion \citep[ hereafter M01]{moraux2001}. Very large
area proper motion surveys \citep{hambly1999,deacon2004} have
obtained on the other hand a significant census of the
stellar-substellar boundary population of the
cluster.\defcitealias{moraux2001}{M01} The coolest object for which
a spectral type has been obtained is the L0 brown dwarf candidate
\object{Roque~25} \citep{martin1998b}, with an estimated
theoretical mass of $\sim$0.035~M$_{\sun}$. A dozen fainter objects
than \object{Roque~25} have been identified in deep surveys
(\citealt{festin1998a,bejar2000,dobbie2002,nagashima2003};
\citealt[ hereafter M03] {moraux2003}; \citealt{schwartz2005}), and some were confirmed by
proper motion \citep{bouy2006}.\defcitealias{moraux2003}{M03}

The mass spectrum $dN/dM=f(M)$ of the Pleiades cluster, where $dN$
stands for the number of objects in the mass range $dM$, is best
fitted by a lognormal function (\citetalias{moraux2003};
\citealt{deacon2004}), and its approximation by a power law
$M^{-\alpha}$ in the low mass range 0.6--0.03~$M_{\sun}$ provides a
spectral index $\alpha$ $\sim$~0.6--0.8 (\citealt{dobbie2002};
\citetalias{moraux2003}). Similar values are found in other much
younger open clusters, as for example \object{$\sigma$ Orionis}
($\sim$2--4~Myr), where $\alpha=$~$0.8\pm0.4$ for
$0.2<$~$M(M_{\sun})<$~$0.013$ \citep{bejar2001}, and \object{IC~348}
($\sim$3~Myr), where $\alpha=$~$0.7\pm0.2$ for
$0.5<M(M_{\sun})<0.035$ \citep{tej2002}. It is important to
determine the behaviour of the $\alpha$ index in the lower mass
domain because it can help to discriminate among brown dwarf
formation mechanisms. For instance, for the formation of
gravitationally unstable cores by turbulent fragmentation
\citep{padoan2004}, the mass spectrum is predicted to be
log-normal, with a peak that appears at smaller masses when the
sonic Mach number and the mean density of the cores are greater.

In this paper we present a 1.8~deg$^2$ deep $J$-band survey of
the Pleiades open cluster aimed to find lower-mass brown dwarfs
($<$0.035~$M_{\sun}$), by comparison with a new analysis of
$I$-band data obtained by \citetalias{bouvier1998}. Our goal is
to identify L dwarf cluster candidates and study the location of
the L dwarf sequence in the colour--magnitude diagram.
Preliminary results have been reported in \citet{bihain2005}.
Because the cluster has a well known age, the properties of
Pleiades L-type brown dwarfs could provide useful constraints to
models predicting the evolution of substellar objects.
Additionally, we can estimate the cluster mass spectrum
of the surveyed area at lower masses.

\section{Observations and data reduction}

\subsection{$J$-band survey} \label{J_survey}

\begin{figure}[pht!]
\resizebox{\hsize}{!}{\includegraphics{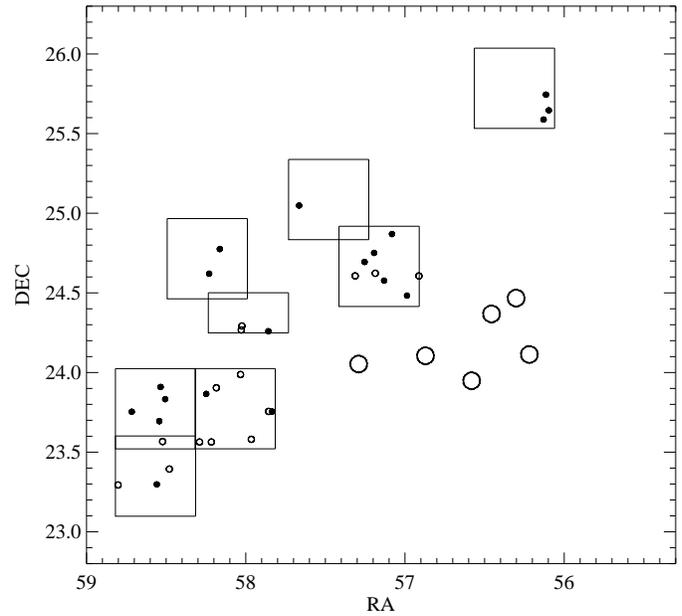}}
\caption{Location of the surveyed Pleiades fields (squares)
relative to the brightest Pleiades members with visual magnitude
$V<$~5 (large open circles). The {\sl small open circles}
represent the brown dwarf candidates from
\citetalias{bouvier1998} with lithium or proper motion
consistent with cluster membership, and the {\sl filled circles}
represent all the other candidates identified in this study.}
\label{surveyfig}
\end{figure}

We obtained $J$-band images on 1998 October 27--28 with the
Omega-Prime ($\Omega'$) instrument at the 3.5~m Telescope of the
Centro Astron\'omico Hispano Alem\'an (CAHA; Calar Alto, Spain).
The fields were specifically chosen to cover the optical $RI$-band
survey from \citetalias{bouvier1998} obtained with the 3.6~m
Canada-France-Hawaii Telescope (CFHT)/UH8K survey. We observed the
fields CFHT 2, 5 (north half), 6, 7, 8, 10, 11 and 13 (see their
spatial distribution in Fig.~\ref{surveyfig}). They correspond to
an area of the cluster of $\sim$1.8 square degrees. The $\Omega'$
camera has a 1024$\times$1024 HgCdTe Hawaii detector with a pixel
scale of 0.40$\arcsec~{\rm pixel}^{-1}$ and a field of view of
$6.8\arcmin\times6.8\arcmin$. To cover each of the
$29\arcmin\times29\arcmin$ UH8K fields (hereafter CFHT fields, as
designed in \citetalias{bouvier1998}) a connecting observation
macro was used. The exposure times and repetitions in each pointing
were $16~{\rm s}\times2$, except for CFHT7 and CFHT8 ($16~{\rm
s}\times3$), and CFHT6 ($18~{\rm s}\times2$). Because the offset
between successive macro pointings corresponded to one third of an
$\Omega'$ field of view, each area was repeated in total 6--9 times
(96--144~s), depending of the CFHT field. The individual images
were sky-subtracted, flat-fielded using dome flats, combined and
aligned in horizontal strips of $\sim30\arcmin\times6.8\arcmin$,
using standard scripts and routines within the {\tt
IRAF\footnote{IRAF is distributed by the National Optical Astronomy
Observatories, which are operated by the Association of
Universities for Research in Astronomy, Inc., under cooperative
agreement with the National Science Foundation.}} environment.
These were later analyzed for aperture and PSF photometry using
{\tt DAOPHOT}. The PSF photometry was calibrated with measurements
from the Two Micron All Sky Survey (2MASS) Point Source Catalog
\citep{cutri2003} and which had errors of less than 0.1~mag. On
average, about five 2MASS calibrators were used per horizontal
strip. The $J$-band completeness (limiting) magnitude was estimated
at the maximum (fainter half maximum) of the histogram of object
counts per magnitude. We found values of $\sim$19.0 and
$\sim$19.5~mag, respectively.

The optical survey in the $RI$ bands from
\citetalias{bouvier1998} was obtained with the UH8K instrument
at the CFHT (Mauna Kea, Hawai'i) on 1996 December 9--13. The
UH8K camera comprises eight Loral 2048$\times$4096 CCD detectors
with a pixel scale of 0.21$\arcsec~{\rm pixel}^{-1}$. These data
were used initially for a study of brown dwarfs and low-mass
stars by \citetalias{bouvier1998}. In the present paper we use
an independent data analysis by \citet{bejar2000}. The
photometric calibration of the CFHT fields was performed by
comparing our photometry to the Cousins photometry of their
candidates and applying an average offset; for the fields CFHT6
and CFHT8 however, different offsets were applied because they
were observed through thin cirrus. The completeness and limiting
magnitudes were $\sim$23.5 and $\sim$25.0~mag in the $R$ band,
and $\sim$22.5 and $\sim$23.5~mag in the $I$ band, respectively.

\begin{table*}
\caption{$RIJ$ photometry and coordinates for the low-mass stars
and brown dwarf candidates.} 
\label{phot_survey_right}
\centering          
     {\scriptsize	
%     {\tiny	
     \begin{tabular}{l l c c c c c}
     	 \hline\hline
	 Name &
	 Names reported$^b$ &
	 $R-I\pm\sigma_{R-I}$ &
	 $I\pm\sigma_{I}$ &
	 $J\pm\sigma_{J}$ &
	 $I-J$ &
	 RA (J2000)\,\,\,\,Dec. (J2000)$^c$
	 \\
	 & & & & & & (${h\,m\,s}$)\,\,\,\,\,\,\,\,\,\,\,\,\,\,(${\degr\,'\,''}$) \\
	 \hline
 \object{BRB 1}    & \object{CFHT-Pl-1}, \object{BPL 242}, \object{MBSC 91}, (*)		    &	  1.76 $\pm$ 0.13     &  16.10 $\pm$ 0.07     &    14.41 $\pm$ 0.03$^d$ &  1.69 &  03 51 51.6  +23 34 50.2 \\
 \object{BRB 2}    & \object{CFHT-Pl-2}, \object{BPL 267}, \object{DH 765}, (*)		    &	  2.00 $\pm$ 0.10     &  16.57 $\pm$ 0.07     &    14.66 $\pm$ 0.06 &  1.91 &  03 52 44.3  +23 54 15.2 \\
 \object{BRB 3}    & \object{CFHT-Pl-6}, (*)				    &	  2.38 $\pm$ 0.10     &  17.13 $\pm$ 0.07     &    14.71 $\pm$ 0.05 &  2.41 &  03 52 07.9  +23 59 14.6 \\
 \object{BRB 4}    & (*) 					    &	  2.38 $\pm$ 0.10     &  17.03 $\pm$ 0.07     &    14.73 $\pm$ 0.04 &  2.30 &  03 44 23.2  +25 38 44.7 \\
 \object{BRB 5}    & \object{CFHT-Pl-3}, \object{HHJ 22}, \object{BPL 272}, \object{MBSC 99}, (*)	    &	  1.94 $\pm$ 0.10     &  16.66 $\pm$ 0.07     &    14.81 $\pm$ 0.06 &  1.85 &  03 52 51.8  +23 33 48.9 \\
 \object{BRB 6}    & \object{CFHT-Pl-4}, \object{BPL 280}, \object{MBSC 101}, (*)		    &	  1.92 $\pm$ 0.10     &  16.85 $\pm$ 0.07     &    14.96 $\pm$ 0.06 &  1.89 &  03 53 09.6  +23 33 48.3 \\
 \object{BRB 7}    & \object{CFHT-Pl-5}, \object{DH 590}, (*)			    &	  2.11 $\pm$ 0.10     &  16.99 $\pm$ 0.07     &    14.98 $\pm$ 0.06 &  2.01 &  03 48 44.7  +24 37 22.7 \\
 \object{BRB 8}    & \object{CFHT-Pl-7}, \object{BPL 253}, \object{MBSC 108}, (*)		    &	  1.84 $\pm$ 0.13     &  17.48 $\pm$ 0.07     &    15.16 $\pm$ 0.07 &  2.32 &  03 52 05.8  +24 17 31.7 \\
 \object{BRB 9}    & \object{CFHT-Pl-12}, \object{BPL 294}, \object{CFHT-PLIZ-6}, (*)	    &	  2.55 $\pm$ 0.10     &  18.00 $\pm$ 0.07     &    15.20 $\pm$ 0.03 &  2.80 &  03 53 55.1  +23 23 37.4 \\
 \object{BRB 10}   & \object{CFHT-Pl-9}, \object{BPL 202}, \object{MHOBD 6}, (*)		    &	  2.17 $\pm$ 0.10     &  17.78 $\pm$ 0.07     &    15.46 $\pm$ 0.07 &  2.32 &  03 49 15.1  +24 36 22.4 \\
 \object{BRB 11}   & \object{CFHT-Pl-13}, \object{Teide 2}, \object{BPL 254}, \object{CFHT-PLIZ-3}, (*)  &	  2.11 $\pm$ 0.13     &  18.14 $\pm$ 0.07     &    15.50 $\pm$ 0.07 &  2.64 &  03 52 06.7  +24 16 01.4 \\
 \object{BRB 12}   & \object{CFHT-Pl-11}, \object{Roque 16}, \object{BPL 152}, (*)  	    &	  2.25 $\pm$ 0.10     &  17.92 $\pm$ 0.07     &    15.64 $\pm$ 0.07 &  2.28 &  03 47 39.0  +24 36 22.1 \\
 \object{BRB 13}   & \object{CFHT-Pl-15}, (*)				    &	  2.44 $\pm$ 0.10     &  18.65 $\pm$ 0.07     &    15.99 $\pm$ 0.07 &  2.66 &  03 55 12.5  +23 17 38.0 \\
 \object{BRB 14}   & \object{CFHT-Pl-21}, \object{Calar 3}, \object{BPL 235}, \object{CFHT-PLIZ-12}, (*) &	  2.49 $\pm$ 0.11     &  18.93 $\pm$ 0.07     &    16.13 $\pm$ 0.05 &  2.80 &  03 51 25.6  +23 45 20.6 \\
 \object{BRB 15}   & \object{CFHT-Pl-25}, \object{BPL 303}, \object{CFHT-PLIZ-20}, (*)	    &	  2.70 $\pm$ 0.13     &  19.68 $\pm$ 0.07     &    16.71 $\pm$ 0.04 &  2.98 &  03 54 05.3  +23 34 00.2 \\
 \object{BRB 16}   & \object{PIZ 1}					    &	  --                  &  20.10 $\pm$ 0.12$^e$ &    16.84 $\pm$ 0.07 &  3.26  &  03 48 31.4  +24 34 37.7 \\
 \object{BRB 17}$^a$   & (**)					    &	  --		      &  20.92 $\pm$ 0.07     &    17.42 $\pm$ 0.06 &  3.50 &  03 54 08.31  +23 54 33.4 \\
 \object{BRB 18}$^a$   & \object{CFHT-PLIZ-28}, (**) 			    &	  2.37 $\pm$ 0.17     &  21.20 $\pm$ 0.07     &    17.61 $\pm$ 0.07 &  3.59 &  03 54 14.08  +23 17 52.2 \\
 \object{BRB 19}$^a$   & --						    &	  2.25 $\pm$ 0.14     &  20.95 $\pm$ 0.09     &    17.79 $\pm$ 0.05 &  3.16 &  03 54 51.49  +23 45 12.2 \\
 \object{BRB 20}$^a$   & \object{CFHT-PLIZ-35}, (*), (**)			    &	  2.55 $\pm$ 0.23     &  21.47 $\pm$ 0.06     &    18.06 $\pm$ 0.07 &  3.40 &  03 52 39.16  +24 46 29.7 \\
 \object{BRB 21}$^a$   & (**)					    &	  2.34 $\pm$ 0.21     &  21.68 $\pm$ 0.08     &    18.14 $\pm$ 0.05 &  3.54 &  03 54 10.27  +23 41 40.3 \\
 \object{BRB 22}$^a$   & \object{\textit{CFHT-PLIZ 2141}}					    &	  2.14 $\pm$ 0.20     &  21.97 $\pm$ 0.08     &    18.31 $\pm$ 0.05 &  3.65 &  03 44 31.27  +25 35 15.1 \\
 \object{BRB 23}$^a$   & (**)					    &	  2.13 $\pm$ 0.27     &  22.03 $\pm$ 0.10     &    18.55 $\pm$ 0.10 &  3.49 &  03 50 39.53  +25 02 54.5 \\
 \object{BRB 24}         & (*)						    &	  2.95 $\pm$ 0.51     &  22.01 $\pm$ 0.09     &    18.71 $\pm$ 0.08 &  3.30 &  03 48 19.65  +24 52 09.5 \\
 \object{BRB 25}         & --						    &	  --		      &  22.24 $\pm$ 0.09     &    18.74 $\pm$ 0.07 &  3.50 &  03 52 59.73  +23 51 56.0 \\
 \object{BRB 26}         & --					    &	  1.93 $\pm$ 0.30     &  22.07 $\pm$ 0.08     &    18.76 $\pm$ 0.10 &  3.31 &  03 48 56.17  +25 09 43.1 \\
 \object{BRB 27}$^a$   & \object{\textit{CFHT-PLIZ 1262}}, (**)					    &	  2.30 $\pm$ 0.32     &  22.64 $\pm$ 0.11     &    18.89 $\pm$ 0.09 &  3.74 &  03 44 27.24  +25 44 41.9 \\
 \object{BRB 28}$^a$   & (*), (**)					    &	  2.64 $\pm$ 0.45     &  22.36 $\pm$ 0.08     &    19.02 $\pm$ 0.10 &  3.33 &  03 52 54.92  +24 37 18.6 \\
 \object{BRB 29}$^a$   & --						    &	  --		      &  22.84 $\pm$ 0.10     &    19.05 $\pm$ 0.07 &  3.79 &  03 54 01.43  +23 49 58.1 \\
 \object{BRB 30}         & --						    &	  1.98 $\pm$ 0.23     &  22.48 $\pm$ 0.10     &    19.17 $\pm$ 0.07 &  3.31 &  03 47 56.96  +24 28 58.4 \\
 \object{BRB 31}         & --						    &	  2.50 $\pm$ 0.25     &  23.05 $\pm$ 0.10     &    19.29 $\pm$ 0.10 &  3.76 &  03 51 25.93  +24 15 32.2 \\
 \object{BRB 32}         & --						    &	  --		      &  22.72 $\pm$ 0.12     &    19.36 $\pm$ 0.09 &  3.36 &  03 49 00.86  +24 41 38.5 \\
 \object{BRB 33}$^a$     & --						    &	  --		      &  23.31 $\pm$ 0.20     &    19.63 $\pm$ 0.09 &  3.68 &  03 51 20.15  +23 45 18.4 \\
 \object{BRB 34}         & --						    &	  --		      &  23.01 $\pm$ 0.16     &    19.64 $\pm$ 0.09 &  3.37 &  03 48 46.55  +24 45 03.2 \\
	\hline					             										 
      \end{tabular}														    
\begin{flushleft}
$^a$ With follow-up observation for proper motion measurement (see Table~\ref{follow-up}).\\

$^b$ (*) \citep{bejar2000}, CFHT-Pl \citepalias{bouvier1998},
BPL \citep{pinfield2000}, MBSC and CFHT-PLIZ
\citepalias{moraux2003}, (**) \citep{bihain2005}, DH
\citep{deacon2004}, \textit{CFHT-PLIZ} \citep{bouy2006}, HHJ~22
\citep{hambly1993}, \object{MHOBD~6} \citep{stauffer1998b},
\object{Teide~2} \citep{martin1998a}, \object{Roque~16}
\citep{zapateroosorio1997b}, \object{Calar~3}
\citep{martin1996}, \object{PIZ~1} \citep{cossburn1997}. In
Fig.~\ref{surveyfig} we overplotted the 14 CFHT-Pl objects (open
circles) and the 20 other objects (filled circles).\\

$^c$ Coordinates of CFHT-Pl, CFHT-PLIZ, \object{PIZ~1} and the
other objects are from \citetalias{bouvier1998},
\citetalias{moraux2003}, \citet{cossburn1997} and the present
study, respectively.\\

$^d$ $J$-band estimate from \citet{martin2000} because our
estimate is bad pixel-contaminated.\\

$^e$ $I$-band estimate corrected for a $\sim0.25$~mag reddening
due to a close star and ray light traces; the offset was
obtained by comparing to objects of similar counts in clean sky
areas of the image. Converting the Kitt~Peak filter measurement
$I_{KP}=19.64$ from \citet{cossburn1997} to Cousins with the
transform equation given by \citet{jameson2002}, we find
$I=19.9$.\\ \end{flushleft}
} 															    
\end{table*} %\end{landscape}

\begin{figure*}[pht!]
\includegraphics[width=16cm]{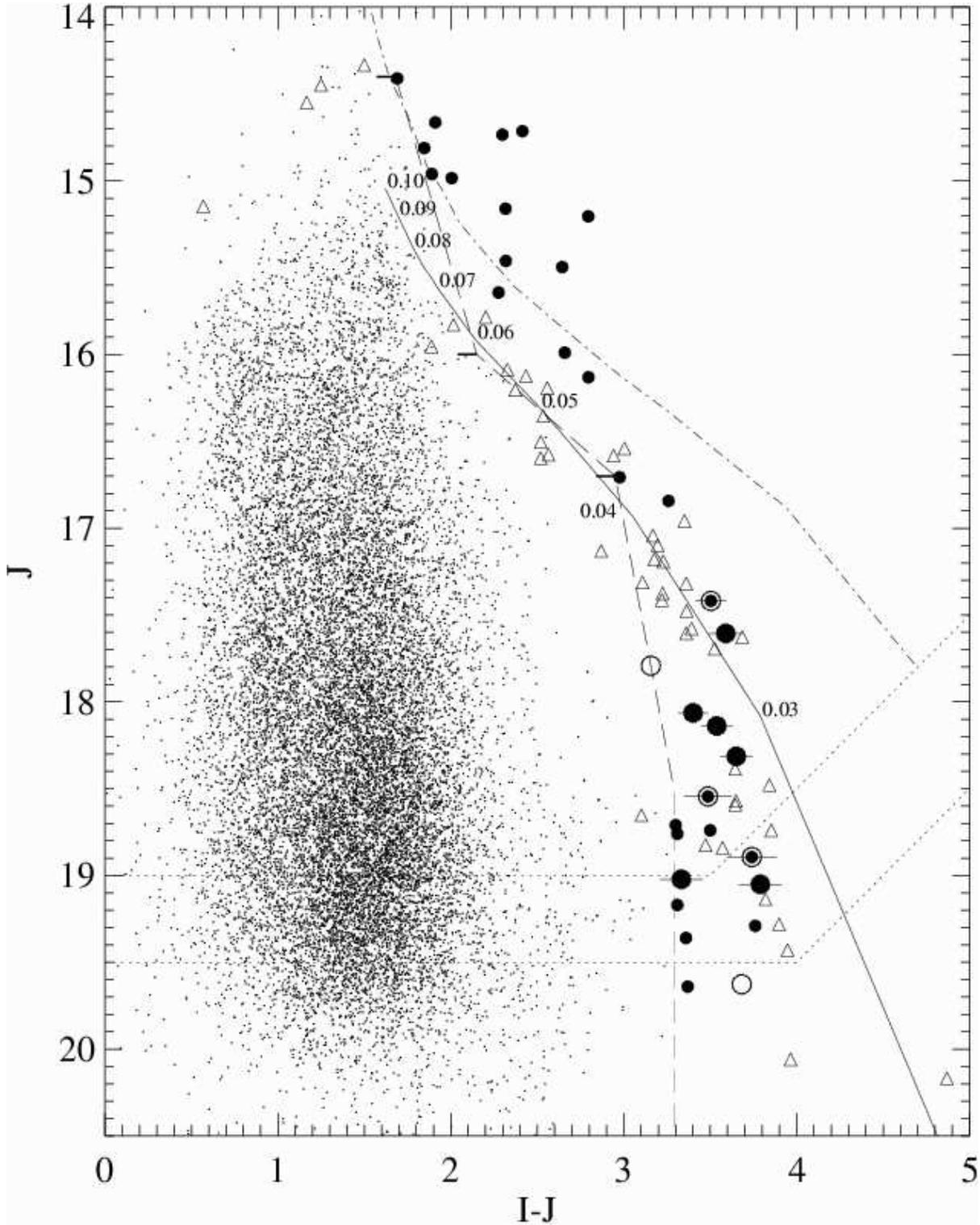}
\caption{$J$~versus~$I-J$ colour--magnitude diagram for the
correlated survey objects. The circular symbols  correspond to
$IJ$ candidate or confirmed cluster members
(Table~\ref{phot_survey_right}), with redder $I-J$ colours than
the dashed line boundary (see text for details).
At $J$ $>17.4$ we plot: proper motion members (large filled
circles), possible proper motion members (circled filled
circles) and non-proper motion members (large empty circles).
The solid line and the dash-dotted line represent the
$\sim$120~Myr DUSTY isochrone \citep{chabrier2000b} and the
$\sim$125~Myr NextGen isochrone \citep{baraffe1998},
respectively, shifted to the distance of the cluster. Masses in
solar units are indicated for the DUSTY isochrone. The triangles
correspond to field dwarfs shifted to the distance of the
cluster. Finally, the upper and lower dotted lines at the bottom
of the diagram indicate the completeness and limiting magnitudes
of the survey, respectively. \label{Jsurvey.can}} \end{figure*}

To identify fainter and redder objects than those found
by~\citetalias{bouvier1998} and~\citet{bejar2000}, we used as
reference the detected $J$-band sources and searched for their
optical counterparts in the $I$-band images using a FORTRAN
correlation program, {\tt CORREL}, kindly provided by
M.~R.~Zapatero Osorio. The correlated objects were plotted in a
$J$~versus~$I-J$ diagram (see Fig.~\ref{Jsurvey.can}) and
contrasted with the 120~Myr DUSTY isochrone from
\citet{chabrier2000b}, shifted to the cluster distance of
133.8$\pm$3.0~pc \citep{percival2005}. As most of our faint
objects were bluer than the theoretical isochrone in the
$J$~versus~$I-J$ diagram, we decided to establish our selection
of candidates based on an empirical criterion. We considered a
bluer envelope to cluster candidates (dashed line in
Fig.~\ref{Jsurvey.can}), defined by the bluest lithium or proper
motion Pleiades objects reported (\citealt{martin2000};
\citetalias{moraux2001}): \object{CFHT-Pl-1}, \object{Calar~1}
and \object{CFHT-Pl-25} (horizontal thick marks in
Fig.~\ref{Jsurvey.can}). The envelope was extended to faint
magnitudes $J>18.5$ at a fixed colour of $I-J=$~3.3, to avoid
most of the M8--9 dwarf contaminants which have bluer colours.
As potential cluster members we have selected 34 objects (named
``BRB") with $J>14.4$ and colours $1.7<I-J<$~3.8
(Table~\ref{phot_survey_right}). We checked that these objects
had colours $R-I\ga1.8$. Some fainter candidates had unreliable
or too faint $R$-band magnitudes and no value is given in
Table~\ref{phot_survey_right}. We discarded as contaminants the
previously identified non-proper motion candidates CFHT-Pl-8
\citepalias{moraux2001} and \object{CFHT-PLIZ-25}
\citepalias{moraux2003}.

Of the 34 BRB candidates, 22 were identified already by previous
surveys (e.g. \citealt{bejar2000}, \citetalias{bouvier1998}; see
footnote of Table~\ref{phot_survey_right} for all the
references). In the magnitude range of the brightest objects we
identify an additional candidate, \object{BRB~4}, which is also
in the 2MASS catalog. This object is not mentioned by
\citetalias{bouvier1998} and is not in the list of proper motion
and $RI$-band photometric candidates from \citet{deacon2004}.
The latter survey overlaps the CFHT survey and its $R$-band
limiting magnitude is greater than the magnitude of
\object{BRB~4}, $R=19.4$. Therefore this object should have been
detected. Possibly it was blended or it is not a proper motion
cluster member.

The magnitude errors of the BRB candidates are 0.06~mag for
$J\sim14.5-17.0$ and 0.08~mag for $J\sim17.0-20.0$. Comparing
our $J$-band magnitudes with those available in the literature
(\citealt{martin2000}; \citealt{pinfield2003}), i.e. those in
the range $J\sim14.5-17.0$, we find:

\begin{enumerate}

\item $<J_{\rm BRB}-J_{\rm Martin}>$ =~$-0.04\pm0.09$,  for the objects in
common \object{CFHT-Pl-2}, 3, 4, 5, 6, 7, 9, 11, 12, 15, 21 and 25.
\object{CFHT-Pl-6} and \object{CFHT-Pl-7} present the greatest differences,
$-$0.18 and $-$0.26~mag, respectively. Both objects lack of H$\alpha$ emission
\citep{martin2000}, but \object{CFHT-Pl-7} has a proper motion consistent with
that of the cluster. In one of our individual $J$-band images, the radial
profile of \object{CFHT-Pl-7} peaks at relatively higher counts and is not well
centred, probably due to a cosmic ray; this might explain the greater
brightness (it is also supported by the difference with 2MASS, $J_{\rm
BRB}-J_{\rm 2MASS}$ =~$-0.14$). For \object{CFHT-Pl-6}, as discussed by
\citetalias{moraux2001}, if this is an equal mass binary as suggested by its
position well above the cluster sequence in the colour--magnitude diagram
(\citetalias{bouvier1998}; see also Fig.~\ref{Jsurvey.can}), the binarity might
affect the determination of its short term proper motion and therefore explain
its deviance from the cluster peculiar motion. The binarity could also explain
some of the magnitude difference observed.

\item $<J_{\rm BRB}-J_{\rm Pinfield}>$ =~$0.04\pm0.09$, for \object{CFHT-Pl-2},
3, 4, 7, 9, 11, 12, 13, 21 and 25. \object{CFHT-Pl-4} and \object{CFHT-Pl-11}
present the greatest differences, 0.19~mag and 0.15~mag, respectively. These
are also found for 2MASS: $J_{\rm BRB}-J_{\rm 2MASS}$ =~0.11 and 0.19,
respectively. But \object{CFHT-Pl-4} and \object{CFHT-Pl-11} have no deviant
proper motions or anomalous positions in the colour--magnitude diagram.

\end{enumerate}

Our $I$-band magnitudes rely on an approximate offset to the magnitudes from
\citetalias{bouvier1998} and have an error of $\sim$0.1~mag. The $I$-band
magnitude from \citetalias{bouvier1998} is obtained after transformation onto
the standard Cousins system, with observation of red Landolt standards. Due to
the lack of very red standard stars, the transformation is linearly
extrapolated for the red and faint objects and could produce systematic errors.
Comparing our magnitudes to those in the Cousins system from
\citetalias{moraux2003}, for the candidates in common and 1--2~mag fainter than
\object{CFHT-Pl-25} (one of the faintest objects from
\citetalias{bouvier1998}), we observe a difference $I_{\rm BRB}-I_{\rm
Moraux}\sim0.1$, which is quite small in comparison with the $I-J>3.3$
threshold that we applied to select the faint candidates.

Finally, from the 34~BRB Pleiades candidates, the 18 faintest have
$J\ga17$ and $I-J\ga3.3$, the magnitudes and colours expected for
L-type dwarfs.

\subsection{Follow-up observations} \label{followup_section}

\subsubsection{$K_{\rm s}$-band imaging} \label{Ksphot}

Follow-up $K_{\rm s}$-band imaging with CAIN-II at the 1.55~m
Telescopio Carlos S\'anchez (TCS; Teide Observatory, Tenerife)
was performed for all 18 $IJ$-band photometric substellar
candidates (see the observation log in Table~\ref{follow-up}).
The CAIN-II camera has a 256$\times$256 HgCdTe Nicmos3 detector
with a pixel scale of 1.00$\arcsec~{\rm pixel}^{-1}$ in wide
field configuration, providing a $4.2\arcmin\times4.2\arcmin$
field of view. The observation macro consisted in 6~s exposures
$\times$ 10 repetitions $\times$ 10 dither positions. Total
exposure times spanned a range of 10--90~min, depending on the
expected brightness of the candidates and the seeing. Data were
sky-subtracted, flat-fielded (using sky flats instead of dome
flats), aligned and combined in a similar way as for the
$J$-band data (Sect.~\ref{J_survey}). We also performed the
photometry and estimated the completeness and limiting
magnitudes as described for the $J$-band. On average, about six
2MASS calibrators were used per field. Average completeness and
limiting magnitudes during the different nights are indicated in
Table~\ref{follow-up}.

\begin{table*}
\protect\caption[]{$HK_{\rm s}$-band follow-up observations.}
\label{follow-up}
\centering          
     {\footnotesize	
     \begin{tabular}{c c c c c}
     	 \hline\hline
	 Telescope/instrument &
	 Filter(s) &
	 Date &
	 Object(s) &
	 $<c>$, $<l>$$^a$
	 \\
	 \hline
         TCS/${\rm CAIN-II}$	& $K_{\rm s}$  &  2004 Dec 7   & \object{BRB 17}, 18, 21, 23, 26 & $\sim$17.6, $\sim$18.2 \\
	 TCS/${\rm CAIN-II}$	& $K_{\rm s}$  &  2004 Dec 8   & \object{BRB 20}, 28            & $\sim$18.1, $\sim$18.6 \\
	 WHT/${\rm LIRIS}$	& $H$          &  2005 Jan 23  & \object{BRB 17}, 23	        & $\sim$19.5, $\sim$20.0 \\
	 WHT/${\rm LIRIS}$	& $H$          &  2005 Jan 24  & \object{BRB 20}, 28	        & $\sim$19.0, $\sim$19.5 \\
	 3.5~m Calar/${\rm \Omega2000}$ & $H$   &  2005 Feb 1   & \object{BRB 17}, 18, 21, \object{Teide 1} & $\sim$18.5, $\sim$19.3 \\
	 TCS/${\rm CAIN-II}$	& $K_{\rm s}$  &  2005 Mar 6   & \object{BRB 25} 		& $\sim$16.5, $\sim$17.0 \\
	 WHT/${\rm LIRIS}$	& $K_{\rm s}$  &  2005 Mar 25  & \object{BRB 27}, \object{Teide 1} & $\sim$18.8, $\sim$19.3 \\
	 TCS/${\rm CAIN-II}$   & $K_{\rm s}$   &  2005 Oct 17  & \object{BRB 19}  		& $\sim$17.5, $\sim$18.5 \\
	 TCS/${\rm CAIN-II}$   & $K_{\rm s}$   &  2005 Oct 18  & \object{BRB 22}  		& $\sim$18.0, $\sim$19.0 \\
	 3.5~m Calar/${\rm \Omega2000}$ & $H$   &  2005 Oct 21  & \object{BRB 19}                & $\sim$19.0, $\sim$19.5 \\
	 TCS/${\rm CAIN-II}$   & $K_{\rm s}$   &  2005 Oct 21  & \object{BRB 24}, 34		& $\sim$17.8, $\sim$18.8 \\
	 3.5~m Calar/${\rm \Omega2000}$ & $H, K_{\rm s}$   &  2005 Oct 22  & \object{BRB 22}     & $\sim$19.0, $\sim$19.5 \\
	 TCS/${\rm CAIN-II}$   & $K_{\rm s}$   &  2005 Oct 22  & \object{BRB 29}  		& $\sim$18.0, $\sim$19.0 \\
	 3.5~m Calar/${\rm \Omega2000}$ & $H$   &  2005 Oct 23  & \object{BRB 29}                & $\sim$18.5, $\sim$19.0 \\
	 TCS/${\rm CAIN-II}$   & $K_{\rm s}$   &  2005 Oct 23  & \object{BRB 31}, 33		& $\sim$18.8, $\sim$19.5 \\
	 3.5~m Calar/${\rm \Omega2000}$ & $K_{\rm s}$   &  2005 Oct 24  & \object{BRB 33}        & $\sim$19.0, $\sim$19.5 \\
	 3.5~m Calar/${\rm \Omega2000}$ & $H$   &  2005 Oct 25  & \object{BRB 33}                & $\sim$19.0, $\sim$19.5 \\
	 TCS/${\rm CAIN-II}$   & $K_{\rm s}$   &  2005 Oct 28  & \object{BRB 30}, 32		& $\sim$18.0, $\sim$19.0 \\
	\hline
      \end{tabular}
\begin{flushleft}
$^a$ Average completeness and limiting magnitudes.\\
\end{flushleft}
      }
\normalsize
\end{table*}

\subsubsection{$HK_{\rm s}$-band imaging for astrometry} \label{HKsphotastr}

For the proper motion determination of the candidates, we
obtained subarcsecond $H$- and $K_{\rm s}$-band images with
the Long-slit Intermediate Resolution Infrared Spectrograph
(LIRIS) at the 4.2~m William Hershel Telescope (WHT; Roque de
los Muchachos Observatory, La Palma) and $H$-band images with
$\Omega2000$ at the 3.5~m Telescope of CAHA (see
Table~\ref{follow-up}).

We decided also to observe the Pleiades brown dwarf
\object{Teide~1} \citep{rebolo1995} with both LIRIS and
$\Omega2000$ to improve its proper motion measurement and to
check if there are any systematic errors in the proper motion
measurements using different instruments. Our first epoch image
was a $12\arcmin\times$~$12\arcmin$ reduced image, obtained with
the TEK3 CCD detector (0.7$\arcsec~{\rm pixel}^{-1}$) at the
prime focus of the 2.5~m Isaac Newton Telescope (INT; Roque de
los Muchachos Observatory, La Palma) on 1995 December 19
(Zapatero Osorio 2005, private communication).

 The LIRIS camera has a 1024$\times$1024 HgCdTe Hawaii detector
with a pixel scale of 0.25$\arcsec~{\rm pixel}^{-1}$ and a
$4.2\arcmin\times4.2\arcmin$ field of view, whereas the
$\Omega2000$ camera has a 2048$\times$2048 HgCdTe Hawaii2
detector with a pixel scale of 0.45$\arcsec~{\rm pixel}^{-1}$
and a field of view of $15.4\arcmin\times15.4\arcmin$. The
observation macros consisted of: 5~s exposures
$\times$~18~repetitions $\times$~5~dithers (LIRIS night 2005
January 23), 5~s exposures $\times$~12~repetitions
$\times$~5~dithers (LIRIS; January 24), 3~s $\times$~20~coadds
$\times$~7~dithers ($\Omega2000$; February 1), 20~s
$\times$~6~repetitions $\times$~5~dithers (LIRIS; March 25), 2~s
$\times$~20~coadds $\times$~15~dithers ($\Omega2000$; October
21, 22 and 23), 2~s $\times$~15~coadds $\times$~20~dithers
($\Omega2000$; October 24) and 3~s $\times$~15~coadds
$\times$~20~dithers ($\Omega2000$; October 25). The total
exposure times were chosen to achieve enough S/N ($\ga$20 in the
peak) for precise measurement of the position of the candidates.
They ranged between 5 and 40 minutes, depending on the expected
brightness of the candidates and the seeing.

The raw images were first bad-pixel corrected using a bad-pixel mask and the
IRAF routine {\tt PROTO.FIXPIX}. The mask was obtained from the flat (sky-flat
for LIRIS and dome-flat for $\Omega2000$) using {\tt
NOAO.IMRED.CCDRED.CCDMASK}. The images were then sky-subtracted -- sky images
were obtained by combining 5--20 consecutive images of similar sky counts --
and divided by the flat.

 The resulting images were distortion corrected, aligned using 10--20 reference
stars and combined. For LIRIS these last three steps were performed with the
{\tt LDEDITHER} task within the package {\tt IRAF.LIRIS.LIMAGE} developed by
J.~A.~Acosta-Pulido. For $\Omega2000$, the images were distortion corrected by
projecting them on the celestial coordinate grid with {\tt MSCIMAGE}. The grid
was obtained by astrometry with the USNO-A2 catalog and the script {\sl
myasrtrom.cl} from E.~Puddu. This script invokes {\tt CCFIND}, that correlates
the catalog sources with those in the image, and {\tt CCMAP}, which computes
the plate solution. Usually, 50--100~stars were accepted by a Legendre order-4
transformation fit and a rejection threshold of 1.5~sigma, providing a
projection transformation with standard deviation $\sim$0.12$\arcsec$ in both
right ascension and declination. The distortion corrected images of
$\Omega2000$ were then aligned and combined using the {\tt INTALIGN} task
within the {\tt LIRIS} package.

The $HK_{\rm s}$-band photometry was obtained similarly as for
the $J$-band data. On average, six and eleven 2MASS calibrators
were used per LIRIS and $\Omega2000$ field, respectively. The
average completeness and limiting magnitudes during the
different nights are indicated in Table~\ref{follow-up}.

\subsection{Proper motion analysis} \label{pm_analysis}

The proper motion measurements were obtained as follows. First we selected
non-saturated and well defined ($S/N>$~$10$) single objects among all those that
appeared within $\sim$3$\arcmin$ of a brown dwarf candidate. We measured their
centres in the first- and the second-epoch images with the IRAF task {\tt
CENTER} ({\tt DIGIPHOT.DAOPHOT} package), with a precision of 0.01--0.1~pixels
in both image dimensions. For \object{BRB~22} and 33 we used the $H$-band
$\Omega2000$ images as second-epoch images, because they provided greater $S/N$
for the candidates than the $\Omega2000$ $K_{\rm s}$-band images. Then we
computed the transformation from second-epoch image to first-epoch image with
{\tt GEOMAP}. For the $\Omega2000$ image of \object{Teide~1}, we performed a
transformation from first- to second-epoch (i.e. from TEK3 to $\Omega2000$),
because the destination image had a wider field and a smaller pixel scale and
therefore provided more accurate equatorial astrometry with the USNO-A2
catalog. With {\tt GEOMAP} we selected the objects whose positions minimized
the transformation error and could thus be considered as reference objects. A
Legendre function of order 3 was used when more than 12~objects were available,
and of order 2 for fewer objects. The error, related to the scatter of these
objects, ranged over 0.03--0.15~pixels. For the candidate \object{BRB~23} the
error in the $y$ dimension was as high as 0.6~pixels, because of vertical
smearing affecting the object positions in the $I$-band image (field CFHT~6).
With {\tt GEOXYTRAN} we used the transformation to predict the candidate
position $(x_{1p},y_{1p})$ in a first-epoch image, and compared it with the
measured one, $(x_{1},y_{1})$, obtaining the pixel shifts
$(\Delta_{x_{1}},\Delta_{y_{1}})=(x_{1p}-x_{1},y_{1p}-y_{1})$. The plate
solution of a first-epoch (or destination) image was obtained with the USNO-A2
catalog and the {\sl myasrtrom.cl} script, in a similar way as explained above.
Then using {\tt WCSCTRAN} we converted the shifts in pixels to shifts in
equatorial coordinates. Due to the precision of the equatorial astrometry
($\sigma\sim0.12\arcsec$), its error contribution to the proper motion was
negligible. But for the LIRIS proper motion measurement of \object{Teide~1},
the best first epoch TEK3 fit had a relatively high standard deviation
($\sigma_{\rm RA}=0.6\arcsec,\sigma_{\rm DEC}=0.7\arcsec$) and the
corresponding proper motion value differed by almost $\sim$1~mas~yr$^{-1}$ with
those derived from less precise fits; thus we added $\sim$1~mas~yr$^{-1}$ to
the proper motion error in this case. In general, we assumed that the proper
motion error was :  {\small \begin{equation} \left(\sigma_{\mu_{\rm \alpha}
{\rm cos} \delta},~\sigma_{\mu_{\rm \delta}}\right)=\left(\frac{f_{\mu_{\rm
\alpha} {\rm cos}
\delta}(\sigma^{2}_{x1}+\sigma^{2}_{x2}+\sigma^{2}_{tx})}{\Delta t},
\frac{f_{\mu_{\rm
\delta}}(\sigma^{2}_{y1}+\sigma^{2}_{y2}+\sigma^{2}_{ty})}{\Delta t}\right),
\end{equation}} \noindent{where $\sigma_{x1,~y1,~x2,~y2}$ are the errors of the $(x,y)$
pixel positions in the first- (1) and second-epoch (2) images, $\sigma_{tx,ty}$
the error of the transformation between these images, $f_{\mu_{\rm \alpha} {\rm
cos} \delta,~\mu_{\rm \delta}}$ the transformation from $(x,y)$ to (RA,~DEC)
coordinates, and $\Delta t$ the time in Julian years between the two epochs of
observation.}

\section{Proper motion Pleiades low-mass brown dwarfs}\label{zresults_dicussion}
\subsection{Photometry and proper motion results}\label{photresults}

The $HK_{\rm s}$ magnitudes for the red and faint brown dwarf
candidates that we have followed-up are given in
Table~\ref{magprop}. The $H$-band magnitude of \object{BRB~17}
is the error-weighted average from the values obtained with
$\Omega2000$ and LIRIS, and the $K_{\rm s}$-band magnitudes of
\object{BRB~22} and 33 are the error-weighted averages from the
values obtained with $\Omega2000$ and TCS. Our 18 follow-up
objects have magnitudes $20.9<I<$~23.3, $17.4<J<$~19.6, and
colours $3.2<I-J<$~3.8, $0.7<J-H<1.3$ and $0.9<J-K_{\rm
s}<$~2.2. Most of them remain good L dwarf candidates, what
supports our $I-J$ selection criterion. But \object{BRB~24}, 26,
30, 31 and 34 present much bluer $J-K_{\rm s}$ colours than
cluster proper motion members with similar $J$-band magnitude
(see Fig.~\ref{JK_li_pm.can}). Moreover, if they were late L or
T dwarfs, they would certainly present much redder $I-J$ colours
($>$4--5~mag). We will assume that these are non-cluster
members. Finally \object{BRB~32}, with its red $J-K_{\rm
s}$ and relatively blue $I-J$ colours, remains a possible
cluster member.

\begin{table*}
\caption{Follow-up $HK_{\rm s}$ photometry of the low-mass brown dwarf candidates.}
     \label{magprop}
\centering
     {\footnotesize	
     \begin{tabular}{l c c c c c c}
     	 \hline\hline
	 Name &
	 $J$ &
	 $I-J\pm\sigma_{I-J}$ &
	 $H\pm\sigma_{H}$ &
	 $J-H\pm\sigma_{J-H}$ &
	 $K_{\rm s}\pm\sigma_{K_{\rm s}}$ &
	 $J-K_{\rm s}\pm\sigma_{J-K_{\rm s}}$
	 \\
	 \hline
\object{BRB 17}$^a$  & 17.42 & 3.50$\pm$0.09 & 16.75$\pm$0.02  & 0.67$\pm$0.06 & 16.20$\pm$0.06 & 1.22$\pm$0.08  \\
\object{BRB 18}$^a$  & 17.61 & 3.59$\pm$0.10 & 16.74$\pm$0.06  & 0.87$\pm$0.09 & 16.08$\pm$0.07 & 1.53$\pm$0.09  \\
\object{BRB 19}$^a$  & 17.79 & 3.16$\pm$0.10 & 16.99$\pm$0.05  & 0.80$\pm$0.07 & 16.67$\pm$0.05 & 1.12$\pm$0.07  \\
\object{BRB 20}$^a$  & 18.06 & 3.40$\pm$0.09 & 17.33$\pm$0.05  & 0.73$\pm$0.08 & 16.56$\pm$0.07 & 1.50$\pm$0.10  \\
\object{BRB 21}$^a$  & 18.14 & 3.54$\pm$0.09 & 17.05$\pm$0.07  & 1.09$\pm$0.09 & 16.39$\pm$0.07 & 1.75$\pm$0.09  \\
\object{BRB 22}$^a$  & 18.31 & 3.65$\pm$0.09 & 17.38$\pm$0.04  & 0.94$\pm$0.06 & 16.69$\pm$0.04 & 1.62$\pm$0.06  \\
\object{BRB 23}$^a$  & 18.55 & 3.49$\pm$0.14 & 17.47$\pm$0.06  & 1.08$\pm$0.11 & 16.65$\pm$0.06 & 1.90$\pm$0.11  \\
\object{BRB 24}      & 18.71 & 3.30$\pm$0.12 &  --	       & --	       & 17.49$\pm$0.07 & 1.22$\pm$0.11  \\
\object{BRB 25}      & 18.74 & 3.50$\pm$0.12 &    --	       &  --	       & 16.91$\pm$0.24 & 1.83$\pm$0.25  \\
\object{BRB 26}      & 18.76 & 3.31$\pm$0.13 &   --	       &  --	       & 17.81$\pm$0.10 & 0.95$\pm$0.14  \\
\object{BRB 27}$^a$  & 18.89 & 3.74$\pm$0.14 &    --	       &  --	       & 17.16$\pm$0.09 & 1.74$\pm$0.12  \\
\object{BRB 28}$^a$  & 19.02 & 3.33$\pm$0.12 & 17.89$\pm$0.05  & 1.13$\pm$0.11 & 17.00$\pm$0.08 & 2.02$\pm$0.13  \\
\object{BRB 29}$^a$  & 19.05 & 3.79$\pm$0.12 & 17.76$\pm$0.05  & 1.29$\pm$0.09 & 16.88$\pm$0.08 & 2.17$\pm$0.11  \\
\object{BRB 30}      & 19.17 & 3.31$\pm$0.12 &    --	       & --	       & 17.65$\pm$0.09 & 1.52$\pm$0.11  \\
\object{BRB 31}      & 19.29 & 3.76$\pm$0.14 &    --	       & --	       & 18.20$\pm$0.15 & 1.09$\pm$0.18  \\
\object{BRB 32}      & 19.36 & 3.36$\pm$0.15 &    --	       & --	       & 17.42$\pm$0.20 & 1.94$\pm$0.22  \\
\object{BRB 33}$^a$  & 19.63 & 3.68$\pm$0.22 & 18.71$\pm$0.07  & 0.92$\pm$0.11 & 17.87$\pm$0.06 & 1.76$\pm$0.11  \\
\object{BRB 34}      & 19.64 & 3.37$\pm$0.18 &    --	       & --			& 18.8$\pm$0.5   &  0.9$\pm$0.5   \\
	\hline
      \end{tabular}
\begin{flushleft}
$^a$ With proper motion measurement.\\
\end{flushleft}
      }
\end{table*}

\begin{table}
\caption{Proper motions of the low-mass brown dwarf candidates and the lithium brown dwarfs \object{Teide~1} and \object{Calar~3}.}
\label{prop}
\centering
     {\small	
     \begin{tabular}{c c c c}
     	 \hline\hline
	 Name &
	 Instrument &
	 $\mu_{\rm \alpha} {\rm cos} \delta \pm \sigma_{\mu_{\rm \alpha} {\rm cos} \delta}$&
	 $\mu_{\rm \delta} \pm \sigma_{\mu_{\rm \delta}}$ \\
	 & & (mas yr$^{-1}$) & (mas yr$^{-1}$) \\
	 \hline
	 \object{Teide 1}  & LIRIS	& 22.5$\pm$3.6 & $-$40.5$\pm$4.0 \\
	 \object{Teide 1}  & $\Omega2000$ & 19.2$\pm$2.5 & $-$45.5$\pm$3.4  \\
	 \object{Calar 3}  & $\Omega2000$ & 24.5$\pm$2.2 & $-$32.6$\pm$2.2  \\
	 \hline
	 \object{BRB 17} & LIRIS        & 20.9$\pm$1.5 & $-$22.8$\pm$2.4  \\
	 \object{BRB 17} & $\Omega2000$ & 17.7$\pm$2.9 & $-$20.7$\pm$3.1  \\
	 \object{BRB 18} & $\Omega2000$ & 28.4$\pm$2.7 & $-$37.4$\pm$2.7 \\
	 \object{BRB 19} & $\Omega2000$ & 75.7$\pm$1.0 & $-$143.9$\pm$7.7  \\
	 \object{BRB 20} & LIRIS	& 21.8$\pm$3.0 & $-$48.1$\pm$2.0  \\
	 \object{BRB 21} & $\Omega2000$ & 18.1$\pm$3.4 & $-$49.1$\pm$4.1  \\
	 \object{BRB 22} & $\Omega2000$ & 18.8$\pm$3.5 & $-$51.9$\pm$2.9  \\
	 \object{BRB 23} & LIRIS	& 29.2$\pm$3.0 & $-$25.6$\pm$14.7 \\
 	 \object{BRB 27} & LIRIS        & 35.3$\pm$3.7 & $-$48.8$\pm$2.0  \\
	 \object{BRB 28} & LIRIS        & 15.1$\pm$3.1 & $-$37.4$\pm$2.1  \\
	 \object{BRB 29} & $\Omega2000$ & 26.4$\pm$3.3 & $-$48.8$\pm$2.9  \\
	 \object{BRB 33} & $\Omega2000$ & $-$4.1$\pm$3.0 & $-$5.2$\pm$3.7  \\
	\hline
      \end{tabular}
      }
\end{table}

\begin{figure}[ht!]
\resizebox{\hsize}{!}{\includegraphics{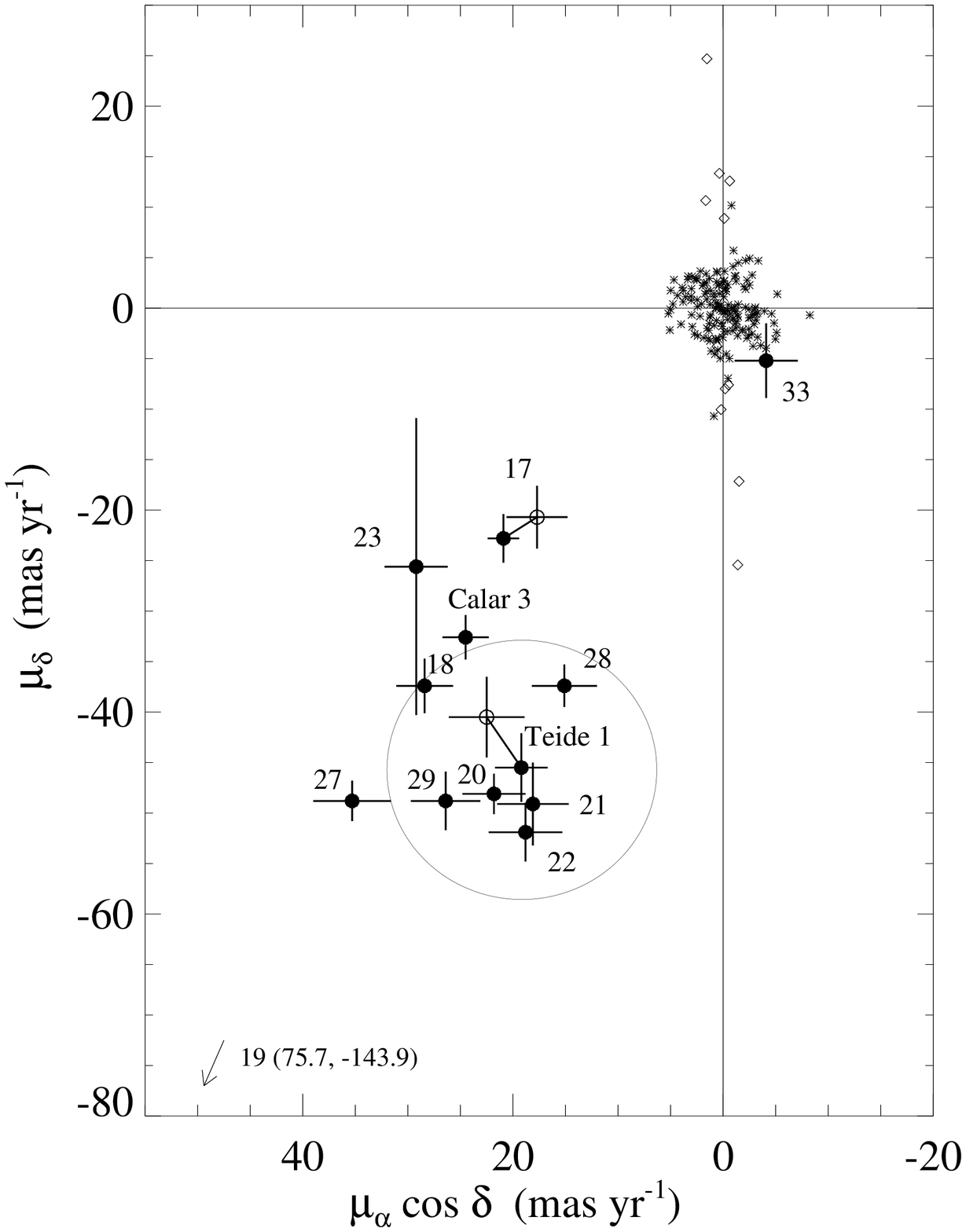}}
\caption{Vector point diagram of proper motion for the candidates
(small circles with BRB number indicated) and the reference objects
(asterisks). The results for the Pleiades lithium brown dwarfs
\object{Teide~1} and \object{Calar~3} are also shown. Double-check
objects are linked by a solid line. The plotted large
circle is centred on the Pleiades average proper motion (see text
for details). The reference objects of \object{BRB~23} (small
squares) present a large dispersion in declination which causes the
large error bar of this candidate. The arrow plotted for
\object{BRB 19} indicates a location outside the diagram.} 
\label{prop_fig} \end{figure}

The proper motion results obtained for 11 of the brown dwarf
candidates and two lithium brown dwarfs (\object{Teide~1} and
\object{Calar~3}) are listed in Table~\ref{prop} and are
represented by small circles in the vector point diagram of
Figure~\ref{prop_fig}. Because \object{Calar~3}
\citep{martin1996,rebolo1996} is at a few arcminutes from
\object{BRB~33}, we could also measure its proper motion and found
a very good agreement with the measurement
of~\citetalias{moraux2001}. We note that the double-check
measurements of \object{Teide~1} and \object{BRB~17} are fully
consistent within the statistic errors. In Fig.~\ref{prop_fig} are
represented also: ($i$) the reference objects used to estimate the
proper motions (asterisks), ($ii$) the circle of radius
three times the greatest component of the average proper motion
error derived from Table~\ref{prop} (($<$~$\sigma_{\mu_{\rm \alpha}
{\rm cos} \delta}$~$>$,~$<\sigma_{\mu_{\rm \delta}}>$)
=~(3.0,~4.3)~mas~yr$^{-1}$). This circle is centred on the cluster
average proper motion, $(\mu_{\rm \alpha} {\rm cos}
\delta,~\mu_{\rm \delta})$
=~(19.15~$\pm$~0.23,~$-$45.72~$\pm$~0.18)~mas~yr$^{-1}$
\citep{robichon1999}, and we will assume that the objects within
are cluster members.

The L dwarf candidates \object{BRB 17}, 23 and 27 lie outside the
circle in Fig.~\ref{prop_fig} and their membership is considered uncertain. The
proper motion dispersion of the seven objects within this circle is
(5.2,~5.9)~mas~yr$^{-1}$. Subtracted quadratically by the average proper motion
error, it yields (4.3,~4.1)~mas~yr$^{-1}$, as an estimate\footnote{At the
cluster distance, it corresponds to a tangential velocity of
$\sim$2.5~km~s$^{-1}$.} of the intrinsic velocity dispersion of Pleiades brown
dwarfs. This velocity dispersion is at least four times that of Pleiades stars
with masses $\ga$1~$M_{\sun}$ (\citealt{jones1970,pinfield1998}) and appears
consistent with the linear relationship of equipartition of energy, between the
velocity dispersion and the inverse square root of the mass for cluster
members, as expected in an nearly relaxed cluster (\citealt{pinfield1998} and
Fig.~4 therein).

We cannot rule out that some of the $IJ$ candidates
located close to the region surrounding the circle in
Fig.~\ref{prop_fig} are in fact cluster members (although field
contaminants are neither excluded, see Sect.~\ref{cont_sect} and
Table~\ref{contam}). This is actually the case of the Pleiades
lithium brown dwarf \object{Calar~3}, suggesting that the
intrinsic velocity dispersion of brown dwarfs could be even
greater than our previous estimate. We need further data (for
instance from lithium observations) to study the membership of
the peculiar $IJ$ candidates. Note also the case of
\object{CFHT-Pl-15} \citepalias{bouvier1998}, which has a proper
motion of (66.5~$\pm$~8.1,~$-$54.1~$\pm$~8.1)~mas~yr$^{-1}$
\citepalias{moraux2001}, thus located far from the cluster point
in the proper motion diagram, although it presents a radial
velocity consistent with that of the cluster, Li absorption,
H$_{\alpha}$ emission and a red $I-K_{\rm s}$ colour 
(\citealt{stauffer1998a};~\citealt{martin2000}).

However, \object{BRB~19} and \object{BRB~33} appear clearly to be
non-proper motion members; \object{BRB~19} has an especially high
proper motion. Finally, our error-weighted average estimate for
\object{Teide~1} is
(20.3~$\pm$~2.1,~$-43.4$~$\pm$~2.6)~mas~yr$^{-1}$, very close to
the cluster proper motion. It also agrees with the measurement
obtained by~\citet{rebolo1995}, the only previous estimate
(16~$\pm$~9,~$-35$~$\pm$~16)~mas~yr$^{-1}$, but with large
conservative errors. These measurements, together with the clear
lithium detection \citep{rebolo1996}, thus fully confirm
\object{Teide~1} as a Pleiades brown dwarf. Also, the photometry
obtained with $\Omega2000$ and LIRIS, $H$ =~$15.54\pm0.06$ and
$K_{\rm s}$ =~$15.07\pm0.09$, respectively, agrees with that of
\citet{jameson2002}, $H$ =~$15.65\pm0.09$ and $K$ =~$15.08\pm0.05$.

\subsection{Pleiades L dwarf sequence} \label{l_seq}

\begin{figure}[ht!] \resizebox{\hsize}{!}{\includegraphics{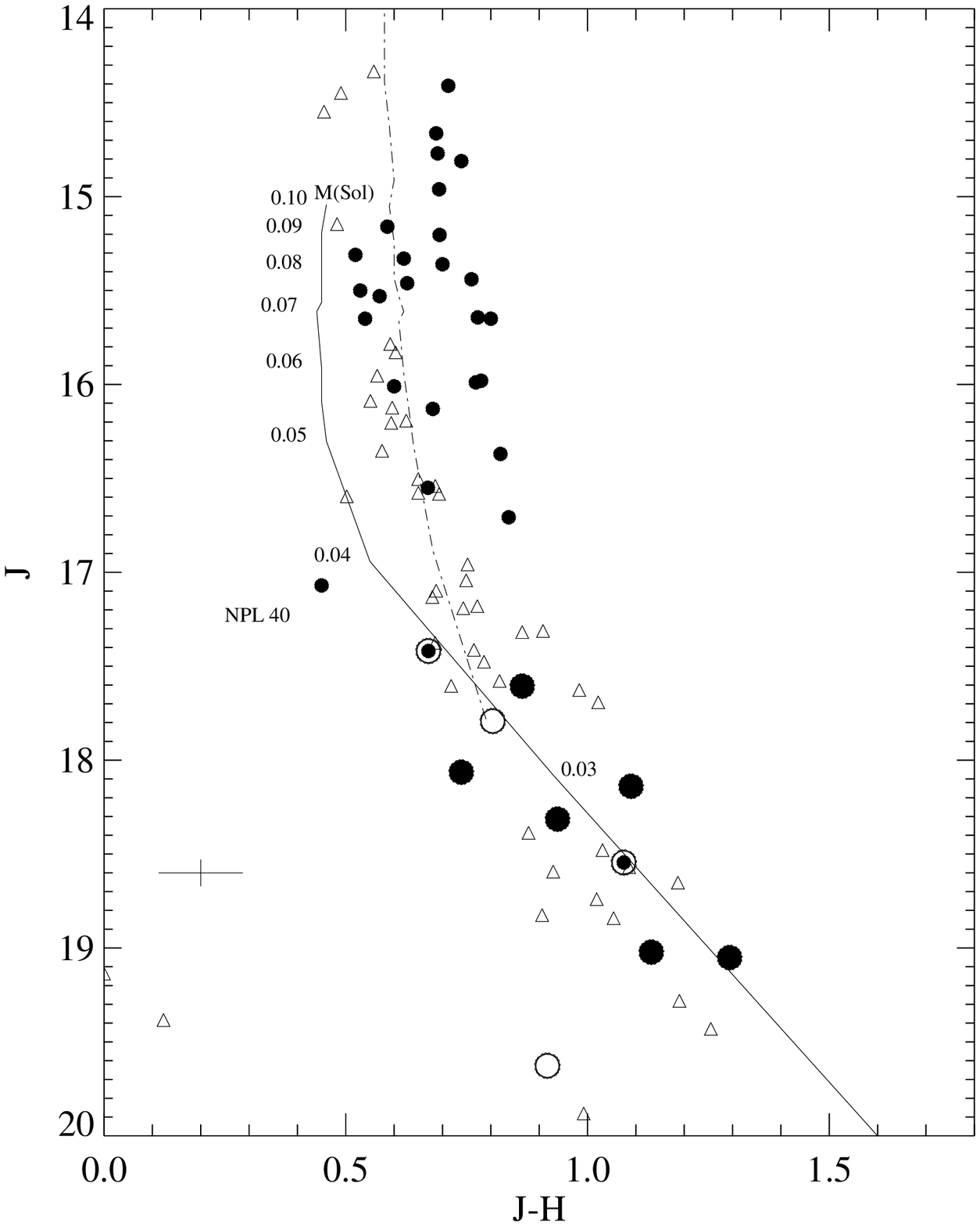}}
\caption{$J$~versus~$J-H$ colour--magnitude diagram for all the Pleiades very
low-mass star and brown dwarf candidates with lithium or proper motion
consistent with membership (filled circles). Their photometry was compiled from
the present study, \citet{martin2000}, the 2MASS catalog,
\citet{zapateroosorio1997b}, \citet{bejar2000} and \citet{jameson2002}.
The circular symbols ($J>17.4$), lines and triangles are defined as in
Figure~\ref{Jsurvey.can}. The typical error bars for the probable and possible
cluster members by proper motion are represented on the left.}
\label{JH_li_pm.can} \end{figure}

\begin{figure}[ht!]
\resizebox{\hsize}{!}{\includegraphics{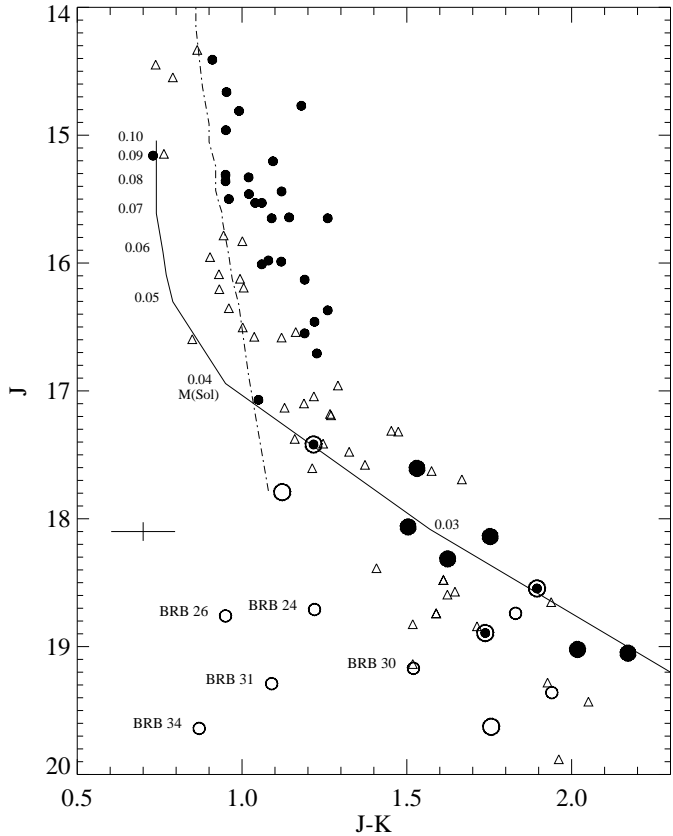}}
\caption{$J$~versus~$J-K$ colour--magnitude diagram, with the same
definition of symbols and error bars as in
Figure~\ref{JH_li_pm.can}. The small empty circles
represent the $IJ$ candidates without proper motion measurements.}
\label{JK_li_pm.can} \end{figure}

In the $J$~versus~$I-J$ colour--magnitude diagram
(Fig.~\ref{Jsurvey.can}), the six cluster proper motion
members (large filled circles) define the Pleiades L dwarf
sequence. Note that the three possible members with relatively high
proper motion (circled filled circles) share the same sequence (as
well as in the near-IR colour--magnitude diagrams,
Figs.~\ref{JH_li_pm.can} and~\ref{JK_li_pm.can}).

In Fig.~\ref{Jsurvey.can} we also present theoretical predictions for the $I$
and $J$ bandpasses. The solid line corresponds to the DUSTY isochrone and the
dash-dotted line corresponds to the NextGen isochrone from \citet{baraffe1998}
for an age 125~Myr. Masses in solar units are shown for the first isochrone.
The NextGen model provides a good fit for $J<$~16.0--16.5 and the DUSTY model
provides a good fit for $J\la$~18. Both models fail to reproduce the
photometric sequence at fainter magnitudes; the former does not account for the
onset of dust formation in cooler atmospheres, whereas the latter overestimates
the $I-J$ colour.

Also in Fig.~\ref{Jsurvey.can} we show the sequence of field dwarfs
(triangles) with parallactic distances, translated to the distance of the
Pleiades cluster. These nearby cool dwarfs are compiled from the literature by
\citet{caballero2006}. Their $I$- and $JHK_{\rm s}$-band photometry are from
\citeauthor{dahn2002} (\citeyear{dahn2000}, 2002) and the 2MASS Point Source
Catalog, respectively. For the M dwarfs, the parallaxes are from
\citeauthor{dahn2002} (\citeyear{dahn2000}, 2002) and the spectral types are
from \citet{golimowski2004}, \citet{leggett2000,leggett2002} and
\citet{dahn2002}. For the L and T dwarfs, the parallaxes are from
\citet{perryman1997}, \citet{dahn2002} and \citet{vrba2004}, and the spectral
types are mostly from \citet{vrba2004}, else from \citet{leggett2002} and
\citet{geballe2001}. In Fig.~\ref{Jsurvey.can} the spectral types range over
$\sim$M3--T2 from top to bottom. At magnitudes $J<$~17 the field dwarfs are
bluer than the cluster brown dwarfs, but at $J>$~17 the field sequence begins
to intercept that of the cluster and to follow it slightly towards redder
colours.

In the $J$~versus~$J-H$ and $J-K$ diagrams (Figs.~\ref{JH_li_pm.can}
and~\ref{JK_li_pm.can}), we show the Pleiades very low-mass stars and brown
dwarf candidates having lithium in their atmospheres or proper motions
consistent with membership in the cluster (filled circles). The circular
symbols ($J>17.4$), lines and triangles are defined as in
Figure~\ref{Jsurvey.can}. The DUSTY and NextGen isochrones provide relatively
good colour predictions for $J>17$ and $J<17$, respectively, in both diagrams.
The fact that the DUSTY model agrees with the observations in the
$J$~versus~$J-H$ and $J-K$ diagrams but not in the $J$~versus~$I-J$ diagram
supports the possible underestimation of the far red flux ($I$ band) relative
to the near-IR flux ($J$-, $H$- and $K$ bands). In Fig.~\ref{JH_li_pm.can}, the
M9 Pleiades brown dwarf \object{NPL~40} \citep{festin1998a,festin1998b} has a
very blue $J-H$ colour, as noted by \citet{pinfield2003}. In
Fig.~\ref{JK_li_pm.can}, the brown dwarf \object{CFHT-Pl-7}, with its magnitude
$J=15.16$ probably enhanced by a cosmic ray, is positioned at the left of the
upper sequence, and the five very blue candidates at $J>$~18.5 (empty circles)
are probable contaminants.

In these near-IR colour--magnitude diagrams, where field dwarfs are
also represented ($\sim$M3--L6), similar sequence overlaps appear
at $J>$~17 as in the $J$~versus~$I-J$ diagram. Thus the sequence of
the L brown dwarf candidates of the Pleiades is nearly
indistinguishable from that of the field dwarfs shifted to the
cluster distance. Note that the average errors in the distances of
the sample field dwarfs are $<\sigma_{\rm d}>$~$=0.4$~pc (M5--M9)
and 1.0~pc (L0--L7), and the average errors in their translated
$J$-band magnitudes, including the photometric errors and the
distance error of 3~pc for the Pleiades cluster, is of only 0.08
and 0.12~mag, respectively. The average error in their $I-J$
colours is 0.03~mag (M5--M9) and 0.06~mag (L0--L7). These magnitude
and colour errors are small in comparison to the (field and
cluster) trends observed.

From the $J$~versus~$I-J$, $J-H$ and $J-K$ diagrams it appears
that, opposite to the Pleiades late M~dwarfs, many of the Pleiades
L~dwarf candidates have the same colours and absolute magnitudes as
their field counterparts. This suggests that they may have the same
spectral energy distributions and luminosities. If the effective
temperatures estimated from the spectral energy distribution do not
depend much on the gravity (as indicated by spectral synthesis),
then, they may have also the same effective temperatures. These
objects would therefore have the same radii. From
\citet{allen2005}, the mean age of field L~dwarfs is in the range
3--4~Gyr and the mean mass is in the range 0.06--0.07~$M_{\sun}$,
which according to the DUSTY models imply radii of
$\sim$0.09~$R_{\sun}$. These objects have almost reached the end of
the contraction. We note however, that according to the same models
(and using the luminosity estimates from Sect.~\ref{subMF}), the
radii of the Pleiades L~brown dwarf candidates are of
$\sim$0.125~$R_{\sun}$, a value inconsistent with the previous one.
Either the predicted evolution of radii with time may have to be
revisited by the models, so the Pleiades L~dwarfs reach radii of
0.09~$R_{\sun}$ in about 120~Myr or, the estimated masses and ages
of the field L~dwarfs may require a revision toward higher masses
or smaller ages. A determination of the radii from the luminosities
and the effective temperatures require precise spectroscopic
observations which are beyond the scope of the present study.

\section{Substellar luminosity function}

\begin{figure}[ht!] 
\resizebox{\hsize}{!}{\includegraphics{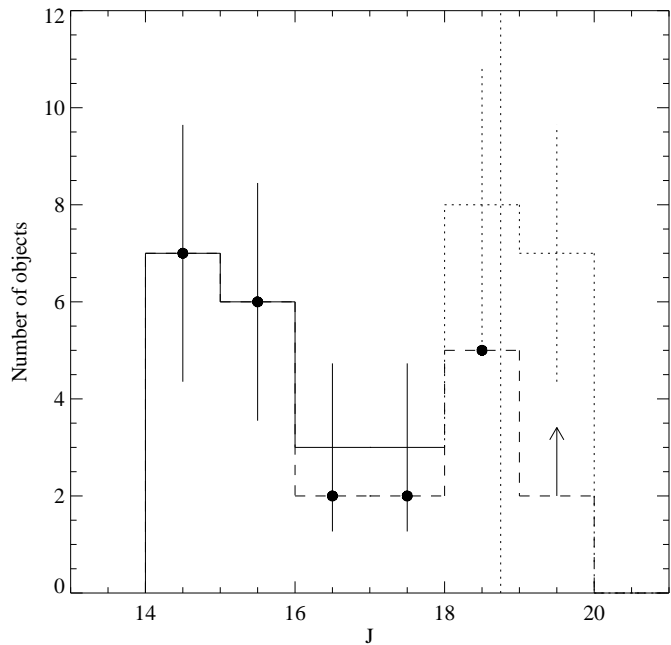}}
\caption{Luminosity function for all the survey candidates
(solid and dotted lines) and after contaminant correction
(dashed line). The vertical dotted line represents the
completeness limit of the survey.} \label{lf_survey}
\end{figure}

The $J$-band luminosity function, the number of objects versus bin of $J$-band
magnitude, is represented in Fig.~\ref{lf_survey} for BRB~1--34, our Pleiades
very low-mass star and brown dwarf $IJ$-band candidates ($J<18$: solid line,
$J>18$: dotted line). The error bars are Poissonian errors, i.e. the square
root of the number of candidates per bin. The vertical dotted line delimits the
completeness of the survey, at $J\sim$~18.75. For this and fainter magnitudes,
the present luminosity function is incomplete (dotted line).

\subsection{Contaminants} \label{cont_sect}

The number of field M and L dwarfs that may contaminate our
sample is obtained as follows. First we subdivide the region in
the $J$~versus~$I-J$ diagram occupied by the 34 objects of
Table~\ref{phot_survey_right} in small rectangular sections.
These sections have heights of half $J$-band magnitude and
widths delimited by the bluer envelope (dashed line in
Fig.~\ref{Jsurvey.can}) and the reddest object.

Then, we consider the typical $I-J$ colours of M- and L-type dwarfs as a
function of the spectral type, based on the same sample of nearby cool dwarfs
as mentioned in Section~\ref{l_seq}. Contaminants in our survey are essentially
M5--L4 dwarfs. Using the relation from \citet{dahn2002} between the spectral
type of M7--L8 dwarfs and the maximum distance at which they can be observed
for a given $J$-band magnitude (a relation that we assume also valid for M5 and
M6 dwarfs), we compute for any spectral type contaminating a given rectangular
section the pair of distances associated to the $J$-band magnitude boundaries
of the section. Each pair of distances defines a volume towards the Pleiades,
e.g. a truncated right pyramid with a square base, with top and base
surfaces subtended by the survey angular area, 1.8~deg$^{2}$. This volume is
located at the average of the pair of distances ($<d>$).

We use the densities of dwarfs in the solar neighbourhood
provided by \citet{kirkpatrick1994} for M5 and M6 spectral types,
and by \citet{cruz2003} for M7--L4 at $d<$~20~pc. We take into
account the completeness and sky coverage of these surveys.
\citet{chen2001} studied the star counts of the Sloan Digital
Sky Survey for several hundred thousands stars at high latitudes
north and south of the Galactic plane. They obtain an
exponential scale height of the old thin disk of $330\pm$~3~pc
(late-type stars) and a Sun's distance from the plane of
$27\pm$~4~pc. We assume that this scale height applies also for
early L dwarfs. To compute the numerical density of a given
spectral type dwarf at an average distance $<d>$ from the Sun,
we multiply the solar neighbourhood density by an exponential
$\exp{-(|<d>\sin(b_{\rm Pl})-z_{\sun}|/z_{\rm h})}$, where
$z_{\rm h}$ is the scale height, $z_{\sun}$ the Sun's distance
to the Galactic plane, $b_{\rm Pl}=$~$-23.52\degr$ the Pleiades
cluster's galactic latitude. 

Finally we multiply the volumes by the corresponding densities and
obtain the statistical numbers of contaminants. Each section in the
$J$~versus~$I-J$ diagram is thus contaminated by field dwarfs of
different spectral types at different distances and with different
densities. In Table~\ref{contam}, we summarize our estimate of
contaminants. In the first column we indicate the $J$-mag
range, in Col. ``$N_{\rm can}$", the number of $IJ$ candidates
(including \object{CFHT-PLIZ-25} and \object{CFHT-PLIZ-25}) and in
Col. ``Confirmed", the BRB numbers of the proper motion and lithium
Pleiades members. In the next three columns, we give the estimated
numbers of contaminants for the spectral type ranges M5--M7, M8--M9
and L0--L4. In column ``total", the total number of contaminants is
provided for the entire spectral type range M5--L4. Nine
contaminants are already found (Col. ``Contaminants found"): (i)
\object{CFHT-Pl-8} \citepalias{moraux2001}, \object{CFHT-PLIZ-25}
\citepalias{moraux2003}, \object{BRB~19} and \object{BRB~33}, (ii)
\object{BRB~24}, 26, 30 31 and 34. The first set of objects is
found by proper motion measurements and the second set by $J-K_{\rm
s}$ photometry (Col. ``Measurement"). The numbers of contaminants
found in each spectral type range are indicated in parenthesis.
Subtracting these from the total number of contaminants, we obtain
the number of remaining contaminants that we may find among our
survey candidates (Col. ``$N_{\rm r.c.}$").

\begin{table*}
\caption{Contaminants among our low-mass star and substellar candidates.}
\label{contam}
\centering          
     {\footnotesize	
     \begin{tabular}{c c c c c c c c c c}
     	 \hline\hline
	 $J$ range &
         $N_{\rm can}$$^{a}$ &
	 Confirmed$^{b}$ &
	 \multicolumn{4}{c}{Estimated numbers of contaminants} &
	 Contaminants &
	 Measurement &
	 $N_{\rm r.c.}$$^{c}$
	 \\
         & & (Li or pm) & M5--M7 & M8--M9 & L0--L4 & total & found & &
	 \\
	 \hline
14.5--15.0   & 7$^{d}$ & 1, 2, 3, 5, 6, 7     & 0.3    & --	& --	 & 0.3 & --	     & --	  & 0 \\
15.0--16.0   & 7       & 8, 9, 10, 11, 12, 13 & 1.6(1) & --	& --	 & 1.7 & \object{CFHT-Pl-18}  & prop. mot. & 1 \\
16.0--17.0   & 3       & 14, 15 	      & 0.8    & 0.3	& 0.1	 & 1.2 & --		    & --	 & 1 \\
17.0--18.0   & 4       & 17, 18 	      & --	     & 1.3(2) & 0.3    & 1.6 & \object{CFHT-PLIZ-25},\object{BRB 19} & prop. mot.	 & 0 \\
18.0--19.0   & 8       & 20, 21, 22, 23, 27   & --	     & 1.7(2) & 1.3    & 3.0 & \object{BRB 24}, 26	    & $J-K_{\rm s}$		 & 1 \\
(19.0--20.0) & 7       & 28, 29 	      & --   & 2.0(2) & 3.8(2) & 5.9 & \object{BRB 30}, 31, 33, 34  & $J-K_{\rm s}$, prop. mot.  & 2 \\
	\hline	          
      \end{tabular}
\begin{flushleft}

$^{a}$ Including \object{CFHT-Pl-18} and \object{CFHT-PLIZ-25}.

$^{b}$ BRB number.

$^{c}$ We assume only 0 and 1 contaminants that remain to be found in the $J$ ranges
15.0--16.0 and 19.0--20.0, respectively, when deriving the effective luminosity
and mass functions (see text, Sect.~\ref{cont_sect}).\\

$^{d}$ Including \object{BRB 1} which has $J=14.41$.\\

\end{flushleft}
      }
\end{table*}

After correcting the number of $IJ$ candidates for contaminants
in each $J$-mag~range, we find an agreement with the number of
confirmed Pleiades members, except for 15.0--16.0 and
19.0--20.0~mag. In these ranges one more contaminant than
observed is predicted. For the latter range, this can be
explained by its incompleteness. In both cases, we assume the
observed number of confirmed members as the effective value when
deriving the effective luminosity and mass functions, thus
decreasing by one the value of $N_{\rm r.c.}$. In
Fig.~\ref{lf_survey}, the effective luminosity function is
represented by the dashed histogram.

\section{Substellar mass function}\label{subMF}

Finally, we present the implications of our results in the
substellar mass function of the cluster surveyed area. The
bolometric luminosity and effective temperature predictions of the
DUSTY and NextGen models are mostly independent of the atmospheric
properties \citep{chabrier2000b}, in contrast to the predicted
magnitudes (e.g. for the M--L transition), which are not. We
decided to infer a bolometric luminosity using a $J$-band
bolometric correction and then to compare it with the model
prediction for the distance and age of the Pleiades to obtain the
mass.

Adopting the Pleiades cluster distance from
\citet{percival2005}, we convert the $J$-band magnitudes to
absolute magnitudes $M_{J}$, and add a bolometric correction
$BC_{J}$ depending on $I-J$. This bolometric correction is
obtained by \citet{caballero2006} from a fit to ($M_{\rm
bol}-M_{J}$,~$I-J$) of the nearby cool dwarfs mentioned in
Section~\ref{l_seq}. Binaries, peculiar dwarfs and dwarfs with
line emission are not taken into account in the fit. For the
nearby M dwarfs, the bolometric magnitudes are from
\citet{golimowski2004}, \citet{leggett2000,leggett2002} and
\citet{dahn2002}. For the nearby L and T dwarfs, these are
mostly from \citet{vrba2004}, else from \citet{leggett2002} and
\citet{geballe2001}. The bolometric magnitudes $M_{\rm
bol}=M_{J}+BC_{J}$ that we obtained are then normalized to the
solar value $M_{\rm bol,\sun}=$~4.74 \citep{livingston2000} and
converted to luminosities. Using these luminosities, we
interpolate linearly the theoretical data points (DUSTY model
for BRB~8--34 and NextGen model for BRB~1--7) and obtain the
masses. These are grouped into five bins: two low-mass stellar
and three substellar bins (see Table~\ref{massbin}). The lowest
mass bin is incomplete ($J\ga18.75$). The differences between
the masses derived from the bolometric luminosities of the
NextGen- and DUSTY models for the fainter objects BRB~8--22,
which extend into the lower luminosity range of the NextGen
model, are smaller than 0.002~$M_{\sun}$ and do not affect the
distribution of objects in these bins. Each bin contains numbers
of BRB candidates ($N_{\rm BRB}$) and contaminants ($N_{\rm
cont}$), identifiable from
Tables~\ref{phot_survey_right}~and~\ref{contam}. The width of
the mass bins within the completeness of the survey is such that
each bin contains approximately the same effective number of
objects, $N_{\rm eff}=N_{\rm BRB}-N_{\rm cont}$.

\begin{table}
\caption{Mass bins for our mass function.}
\label{massbin}
\centering
     {\small	
     \begin{tabular}{c c c c c}
     	 \hline\hline
	 $\Delta J$       &
	 $\Delta M$       &
	 $N_{\rm BRB}$    &
	 $N_{\rm cont}$ &
	 $N_{\rm eff}$           \\
	  &($M_{\sun}$) & & & \\
	 \hline
	 14.41-14.97  & 0.147-0.104 &  6 &  0 &  6  \\
	 14.97-15.57  & 0.104-0.075 &  5 &  0 &  5  \\
	 15.57-17.51  & 0.075-0.039 &  6 &  1 &  5  \\
	 17.51-18.75  & 0.039-0.026 &  8 &  3 &  5  \\
	 (18.75-19.65)  & 0.026-0.018 &  9 &  6 &  3  \\
	\hline
      \end{tabular}
      }
\end{table}

\begin{figure}[ht!]
\resizebox{\hsize}{!}{\includegraphics{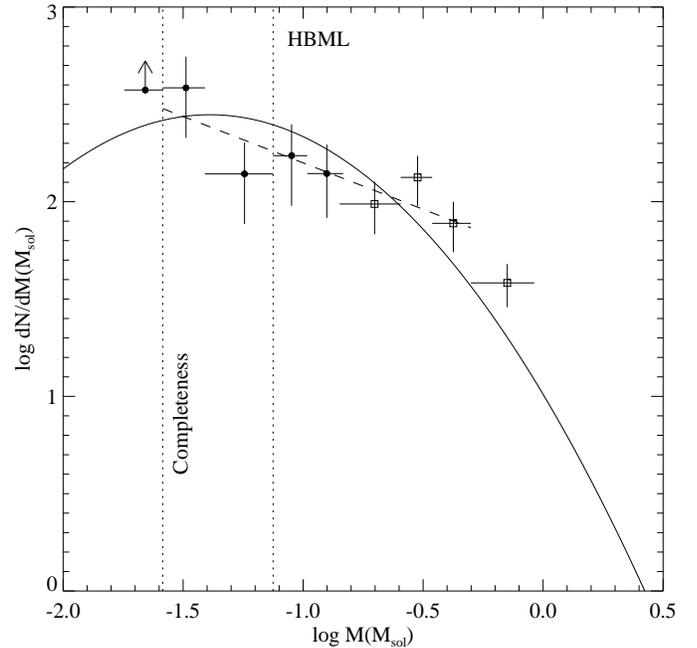}}

\caption{The derived mass function for the Pleiades survey
area. The filled circles are for the BRB objects, whereas the
squares are for the objects from \citet{deacon2004}. The right
dotted line indicates the hydrogen-burning mass limit. The lower
limit arrow for the lowest mass bin indicates that this bin is
incomplete. The curve represents the log normal function obtained
by \citet{deacon2004} scaled down to the total number of objects in
the survey area and between 0.5~$M_{\sun}$ and the completeness
mass limit 0.026~$M_{\sun}$ (56). The dashed line represents the
power law fit to the data points between 0.5 and 0.026~$M_{\sun}$,
with $\alpha=0.5\pm0.2$.}

\label{mf_fig}
\end{figure}

In Fig.~\ref{mf_fig}, we show our results in the mass spectrum
representation, $\log{dN/dM}$,~$\log{M}$. Error bars refer to the
Poissonian error. We show also the bins for the $\sim$50~massive
proper motion Pleiades objects from \citet{deacon2004} which are
present in our fields\footnote{The objects in common with our
sample, \object{DH 765} and \object{DH 590}, are taken into account
only once and in our mass spectrum bins.}. For the $\sim$35 fainter
objects, corresponding essentially to M dwarfs ($T_{\rm eff} \la
3600$~K), we obtain the masses with the method explained above, but
using $I$- and $J$-band magnitudes from \citet{deacon2004} and
2MASS, respectively, because these objects are saturated in our
images. For the brighter ones we proceed similarly but, this time,
using a synthetic $K$-band bolometric correction, function of the
$(J-K)_{\rm CIT}$~colour and determined by \citet{houdashelt2000}
for FGK stars with 4000~K$~\le T_{\rm eff} \le$~6500~K. We assume
solar metallicity and $\log g=4$. The CIT colour is transformed
into the 2MASS equivalent (see the Explanatory Supplement to the
2MASS All Sky Data
Release\footnote{http://www.ipac.caltech.edu/2mass/releases/allsky/doc/explsup.html},
Cutri et al. 2006) and the bolometric correction is applied on
2MASS $JK_{\rm s}$-band photometry. For the objects in the K--M
transition, the masses obtained by any of these two methods are
within the highest mass bin ($M>0.5$~$M_{\sun}$), thus fixing the
number of objects in this bin. A linear fit to all the data points
between 0.5 and 0.026~$M_{\sun}$ provides an $\alpha=0.5\pm0.2$ for
the power law $dN/dM \propto$~$M^{-\alpha}$. It is compatible with
the mass bin beyond the completeness limit of the survey (lower
limit arrow at the left of the vertical dotted line,
Fig.~\ref{mf_fig}), defined by the three lowest mass possible or
probable cluster members by proper motion, \object{BRB~27}, 28 and
29. Our estimate of $\alpha$ is in agreement with that from
\citetalias{moraux2003}, $0.60\pm0.11$, and signifies that the
number of objects per unit mass still increases at these low
masses. In Fig.~\ref{mf_fig} the lognormal function obtained by
\citet{deacon2004} is also represented, scaled down so that its
integral over the mass range 0.5--0.026~$M_{\sun}$ is equal to the
number of objects in the survey area and in this mass range (56).

Our estimate of $\alpha$ is similar to those for younger clusters as
\object{$\alpha$~Per} ($\sim$80~Myr), \object{IC~4665} (30--100~Myr) and
\object{$\lambda$~Orionis} ($\sim$5~Myr). Their values are: 0.56
($0.2>M(M_{\sun})>$~0.06; \citealt{barradoynavascues2002}), 0.6
($1>M(M_{\sun})>$~0.04; \citealt{dewit2006}) and 0.60~$\pm$~0.06
($0.86>M(M_{\sun})>$~0.024; \citealt{barradoynavascues2004}), respectively. The
presence of low-mass brown dwarfs after more than 100~Myr in the
\object{Pleiades} open cluster, in numbers relative to stars which are similar
to those found in other younger clusters, implies that differential evaporation
of low-mass members relative to more massive ones has not been very significant
-- in contrast to what may occur in the $\sim$600~Myr~old \object{Praesepe}
open cluster \citep{chappelle2005}. Comparison of observations to evolutionary
models of the Pleiades open cluster suggests indeed that only marginal
differential evaporation of the massive brown dwarfs has occurred
\citep{moraux2004,moraux2005}. Therefore searches for the lowest mass brown
dwarfs and even planetary-mass objects should be conducted. A Pleiades giant
planetary-mass object at the deuterium-burning mass limit has a predicted
effective temperature of $\sim$1300~K (DUSTY and COND models). This effective
temperature corresponds to late L and early T field dwarfs
\citep{golimowski2004} and magnitudes $J>$~20 \citep{knapp2004} when scaled to
the distance of the cluster. Such a free-floating object in the Pleiades open
cluster may be detected in a photometric search with mid-size telescopes
($\sim$4~m) but spectroscopic classification will require a very large
telescope of 8--10~m~class. The discovery of such low-mass Pleiades objects
will permit to compare with free-floating low-gravity candidates in the solar
neighbourhood, as for example the T6 brown dwarf \object{SDSS
J111010.01+011613.1} \citep{knapp2004}. Extrapolating the mass spectrum slope
to lower masses, we find that our survey area may contain five brown dwarfs
with masses 0.026--0.013~$M_{\sun}$ and four planetary-mass objects with masses
0.013--0.05~$M_{\sun}$ (or two per square degree). In the case where the mass
spectrum is log normal down to planetary masses, these values drop to three and
one, respectively (or one planetary-mass object per square degree).

\section{Conclusions} \label{mf_section}

As a result of an $IJ$-band survey in the Pleiades, we have
found 18 brown dwarf candidates, most likely of L-type. They
have 17.4~$<J<$~19.7, $I-J>$~3.2 and theoretical masses
$\sim$0.040--0.020~$M_{\sun}$. The near-IR follow-up for
proper motion of 11 candidates permits us to confirm six as
cluster members. Another three remain as possible members. Two
are clear non-members. The near-IR photometric follow-up of the
seven other candidates indicates that at least five are probable
contaminants, because their $J-K_{\rm s}$ colours are bluer than
those of the cluster proper motion members. These latter
determine the L brown dwarf photometric sequence of the
Pleiades. The sequence in the $J$~versus~$I-J$ colour--magnitude
diagram at $J>18$ is bluer than that predicted by the DUSTY
atmospheres model, whereas it agrees with the predictions in the
$J$~versus~$J-H$ and $J-K$ diagrams. Moreover the sequence
overlaps that of the field L~dwarfs shifted to the cluster
distance in the three colour--magnitude diagrams, suggesting
that the Pleiades and field L~dwarfs may have comparable
spectral energy distributions and luminosities, and thus
possibly similar radii.

We find evidence for an intrinsic velocity dispersion of
Pleiades brown dwarfs at least four times that of Pleiades stars
more massive than the Sun. The estimated value $>$4~mas~yr$^{-1}$
appears consistent with the kinematical expectations for brown
dwarfs in a nearly relaxed cluster.

Correcting for contaminants and comparing with the proper motion
objects from \citet{deacon2004} in the same fields, we obtain a
substellar mass spectrum for the surveyed area of the
cluster. For a power law $dN/dM \propto$~$M^{-\alpha}$ fit in the
mass range 0.5--0.026~$M_{\sun}$, we find a spectral index
$\alpha=$~0.5~$\pm$~0.2 and in agreement with previous results. The
slope is similar to that of much younger open clusters, suggesting
the absence of significant differential evaporation of the low-mass
brown dwarfs in the Pleiades. This supports deeper searches to
detect lower mass free-floating objects in the cluster.

\begin{acknowledgements}

The 3.5~m Telescope is operated jointly by the Max-Planck Institut
f{\" u}r Astronomie and the Instituto de Astrof{\'i}sica de
Andaluc{\'i}a (CSIC) at the Centro Astron{\'o}mico Hispano
Alem{\'a}n (CAHA) at Calar Alto. The Canada-France-Hawaii Telescope
(CFHT) is operated by the National Research Council of Canada, the
Institut National des Sciences de l'Univers of the Centre National
de la Recherche Scientifique of France, and the University of
Hawaii. The William Hershel Telescope (WHT) is operated on the
island of La Palma by the Isaac Newton Group in the Spanish
Observatorio del Roque de los Muchachos of the Instituto de
Astrof{\'i}sica de Canarias. The Carlos S{\'a}nchez Telescope (TCS)
is operated by the Instituto de Astrof{\'i}sica de Canarias at the
Teide Observatory, Tenerife. This publication makes use of data
products from the Two Micron All Sky Survey, which is a joint
project of the University of Massachusetts and the Infrared
Processing and Analysis Center/California Institute of Technology,
funded by the National Aeronautics and Space Administration and the
National Science Foundation. We would like to thank J.~Bouvier for
allowing us to use the $RI$-band data and for refereeing this
paper. We thank Terry Mahonay for the English corrections. The
LIRIS observations were obtained with the assistance and help of
the LIRIS commissioning team. We thank J.~I.~Gonz\'alez Hern\'andez
and I.~Villo for obtaining near-IR data for us, during LIRIS night
2005 January 24 and $\Omega2000$ night 2005 February 1,
respectively. We thank also N.~Deacon, H.~Bouy, E.~Mart{\' i}n,
M.~R.~Zapatero Osorio and E. Delgado-Donate for valuable
discussions. J.A.P \& A.M. acknowledge the Plan Nacional de
Astronomia y Astrofisica (AYA2004-03136), which supported part of
this work.

\end{acknowledgements}

\bibliographystyle{aa}
\bibliography{astronomy_ref}
\end{document}